\documentclass[aps, prl, showpacs, twocolumn, amsfonts, amsmath, amssymb, superscriptaddress, floatfix]{revtex4}

\usepackage[T1]{fontenc}
\usepackage[latin9]{inputenc}
\usepackage{color}
\usepackage{graphicx}

\def\be{\begin{equation}}
\def\ee{\end{equation}}
\def\bea{\begin{eqnarray}}
\def\eea{\end{eqnarray}}

\begin{document}

\title{Lieb-Liniger model: emergence of dark solitons in the course of measurements of particle positions}

\author{Andrzej Syrwid} 
\affiliation{
Instytut Fizyki imienia Mariana Smoluchowskiego, 
Uniwersytet Jagiello\'nski, ulica Profesora Stanis\l{}awa \L{}ojasiewicza 11 PL-30-348 Krak\'ow, Poland}
 
\author{Krzysztof Sacha} 
\affiliation{
Instytut Fizyki imienia Mariana Smoluchowskiego, 
Uniwersytet Jagiello\'nski, ulica Profesora Stanis\l{}awa \L{}ojasiewicza 11 PL-30-348 Krak\'ow, Poland}
\affiliation{Mark Kac Complex Systems Research Center, Uniwersytet Jagiello\'nski, ulica Profesora Stanis\l{}awa \L{}ojasiewicza 11 PL-30-348 Krak\'ow, Poland
}

\pacs{03.75.Hh, 03.75.Lm}

\begin{abstract}
Lieb-Liniger model describes bosons with contact interactions in one-dimensional space. In the limit of weak repulsive particle interactions, there are two types of low lying excitation spectrum. The first is reproduced by the Bogoliubov dispersion relation, the other is believed to correspond to dark soliton excitations. While there are various evidences that the type II spectrum is related to dark solitons, it has not been shown that measurements of positions of particles reveal dark soliton density profiles. Here, we employ the Bethe ansatz approach and show that dark solitons emerge in the measurement process if the system is prepared in an eigenstate corresponding to the type II spectrum. We analyze single and double dark solitons as well as weak and strong interaction regime.
\end{abstract}

\maketitle

Solitons are solutions of non-linear wave equations that do not change their shapes in the time evolution \cite{KivsharOpticalSol}. Solitons are intensively studied in non-linear optics but they can also be created by matter waves. Ultra-cold bosonic atoms form a Bose-Einstein condensate (BEC) where all particles occupy the same single particle wave function. The latter is accurately described by the Gross-Pitaevskii equation (GPE) \cite{pethicksmith}. Such a single-particle non-linear equation possesses bright and dark soliton solutions in one-dimensional (1D) space for attractive and repulsive particle interactions, respectively \cite{Zakharov71,Zakharov73}. Both kinds of the solitons have been realized experimentally \cite{burger1999,denschlag2000,strecker2002,khaykovich2002,becker}. There are also attempts to create dark solitons in superfluid Fermi systems 
\cite{yefsah2013,Ku2014,sacha2014}. 

Solitons constitute broken-symmetry states of many-body atomic systems. In the absence of an external trapping potential, a many-body Hamiltonian is invariant under spatial translation of all particles by the same vector. Consequently system eigenstates are also eigenstates of the corresponding spatial translation operator and a plot of the reduced single particle probability density cannot reveal any solitonic profile because it is uniform in space. However, even if a system is initially prepared in an eigenstate, solitons can emerge due to symmetry-breaking perturbation, e.g. measurements or atomic losses. This phenomenon is relatively easy to analyze for attractive particle interactions where the bright soliton formation is the result of a symmetry breaking in the system prepared initially in the ground state \cite{Lai89,Lai89a,castinleshouches,Weiss09,delande2013}. On the contrary, dark solitons are related to collectively excited states of a repulsively interacting atomic gas. The question if there exist many-body eigenstates which are able to reveal dark solitons in the course of measurements of particle positions is highly non-trivial \cite{corney97,corney01,martin2010b,delande2014,kronke15}. 

The Lieb-Liniger model, that describes bosons in 1D space with contact interactions, can be solved by means of the Bethe ansatz \cite{Lieb63,Lieb63a,Korepin93}. In the limit of weak repulsive particle interactions, there are two branches of low lying excitation spectrum \cite{Lieb63a}. The type~I spectrum is reproduced by the Bogoliubov dispersion relation while the type~II spectrum, related to the so-called one-hole excitations, is believed to correspond to dark soliton excitations due to the coincidence with the soliton dispersion relation \cite{kulish76,ishikawa80,komineas02,jackson02,kanamoto08,kanamoto10,karpiuk12,karpiuk15}. Recently Sato {\it et al.} \cite{sato12,sato12a} have shown that superpositions of the type~II eigenstates result in states for which the reduced single particle probability densities reveal shapes very similar to dark soliton profiles. These states are not stationary and the soliton profiles diminish in the time evolution. In the present publication we concentrate on single eigenstates and analyze the long-standing problem whether dark solitons emerge in the course of measurements of particles positions if the system is prepared in a type~II eigenstate.

We consider $N$ bosons with repulsive contact interactions in a 1D space with periodic boundary conditions. The second quantized version of the system Hamiltonian reads,
\be
H=\int_0^Ldx\left[\partial_x\hat\psi^\dagger\partial_x\hat\psi+c\hat\psi^\dagger\hat\psi^\dagger\hat\psi\hat\psi\right],
\label{h}
\ee
where $\hat\psi$ is the bosonic field operator, $L$ is the system size and $c>0$ determines strength of the repulsive interactions. We use units where $2m=\hbar=1$, with $m$ the particle mass. The system is characterized by a parameter $\gamma=\frac{c}{n}$ \cite{Lieb63}, where $n=\frac{N}{L}$ is the particle density. For $\gamma\ll 1$ one deals with the weak interaction limit. The Bethe ansatz allows us to find eigenvalues and eigenvectors, $|\{k\}_N\rangle$, of the Hamiltonian (\ref{h}) which are determined by $N$ parameters, $\{k\}_N=\{k_1,\dots,k_N\}$, that satisfy the Bethe equations $k_jL+2\sum_{m\ne j}^N\arctan\left(\frac{k_j-k_m}{c}\right)=2\pi I_j$. The $I_j$'s are arbitrary integers (half integers) for odd (even) $N$ which unambiguously determine an eigenstate \cite{Korepin93,sato12}. The sets $\{k\}_N$ are often called the sets of quasimomenta because eigenvalues of the total momentum $P(\{k\}_N)=\sum_{j=1}^Nk_j=\frac{2\pi}{L}\sum_{j=1}^NI_j$ and the energy eigenvalues $E(\{k\}_N)=\sum_{j=1}^Nk_j^2$. The ground state of the system corresponds to the ordered set $\{I\}_N=\{-\frac{N-1}{2},-\frac{N-1}{2}+1,\dots,\frac{N-1}{2}\}$. The type~II excitation spectrum is obtained by removing one of $I_j$ and adding $-\frac{N-1}{2}-1$ or $\frac{N-1}{2}+1$ \cite{Lieb63a,sato12,kanamoto10}. This can be called one-hole excitation because one of the consecutive integers (or half integers for even $N$) becomes missing and a new integer appears at the beginning or at the end of the initial sequence $\{I\}_N$. 

{\it Single soliton.}
Let us consider even $N$. The ground state of the system corresponds to the total momentum $P=0$. If in the sequence $\{I\}_N$, we remove $\frac12$ and add $\frac{N+1}{2}$, we obtain the type~II eigenstate, which we denote by $|sol\rangle$, corresponding to the total momentum $P_{sol}=\frac{N\pi}{L}$. The reduced single particle probability density is uniform, i.e. $\rho_1(x_1)=\frac{1}{N}\langle sol|\hat\psi^\dagger(x_1)\hat\psi(x_1)|sol\rangle=\frac{1}{L}$. Thus, the result of the measurement of the position of a single particle is equally probable in the entire space. Suppose that we have performed such a measurement and then ask about the probability density $\rho_2(x_2)$ for a choice of the position $x_2$ of the second particle. More generally we need the conditional probability density for a choice of the position of the $j$-th particle provided $(j-1)$ particles have been already measured \cite{javanainen96,dziarmaga06}. Such a probability density reads
\be
\rho_j(x_j)=\sum_{\{k\}_{N-j}}\frac{|\Gamma_j(x_j,\{k\}_{N-j})|^2}{N-j+1},
\label{rhoeq}
\ee
where the sum runs over all sets of $\{k\}_{N-j}$, i.e. over all eigenstates of the system with the total number of particles equal $N-j$, and 
\bea
\Gamma_j(x_j,\{k\}_{N-j})&=&\sum_{\{q\}_{N-j+1}}\Gamma_{j-1}(x_{j-1},\{q\}_{N-j+1}) \cr
& \times & e^{i[P(\{q\}_{N-j+1})-P(\{k\}_{N-j})]x_j} \cr
& \times & \langle\{k\}_{N-j}|\hat\psi(0)|\{q\}_{N-j+1}\rangle,
\label{gam}
\eea
with 
\bea
\Gamma_1(x_1,\{q\}_{N-1}))&=&e^{i[P_{0}-P(\{q\}_{N-1})]x_1} \cr
& \times & \langle\{q\}_{N-1}|\hat\psi(0)|\psi_0\rangle,
\label{gam1}
\eea 
where $|\psi_0\rangle$ is an initial eigenstate and $P_0$ the corresponding eigenvalue of the total momentum. We will start with $|\psi_0\rangle=|sol\rangle$.
The last terms in (\ref{gam}) and (\ref{gam1}), so called form factors of the field operator, can be easily numerically calculated by means of the determinant formulas \cite{kojima97,caux07} \footnote{Here we assume that any eigenstate denoted by $|\{k\}_{N}\rangle$ or $|\{q\}_{N-j}\rangle$ is normalized to unity which is a different notation as compared to Ref.~\cite{caux07}.}.

\begin{figure}
\includegraphics[width=0.9\columnwidth]{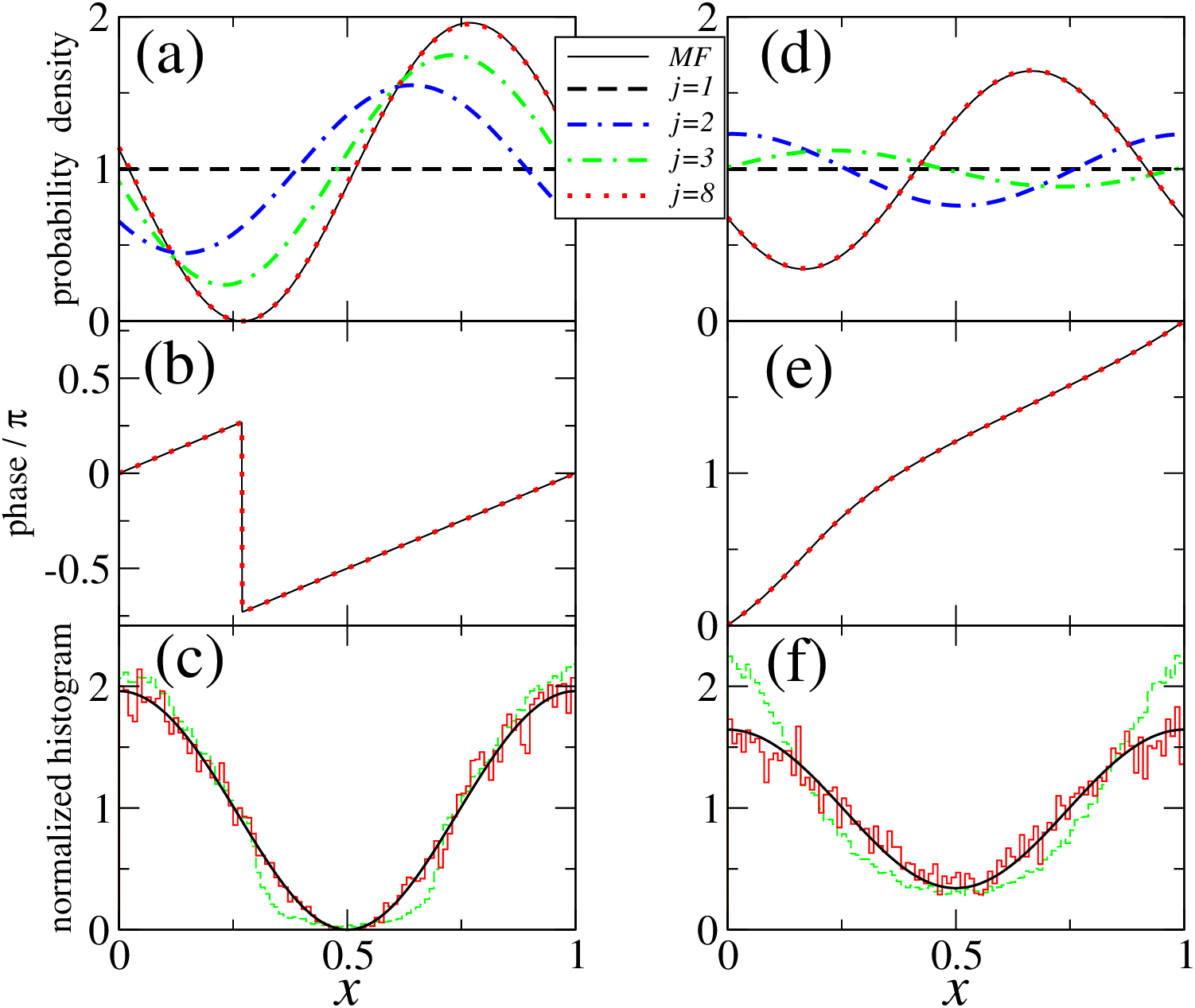}
\caption{(color online) (a)-(c) Results of the particle positions measurements in the system prepared in the eigenstate corresponding to the one-hole excitation with the total momentum per particle $P/N=\frac{\pi}{L}$, for $N=8$, $c=0.08$ and $L=1$, i.e. $\gamma=0.01$. (a) shows an example how the conditional probability density $\rho_j(x)$ (with values of $j$ indicated in the panel) changes after successive measurements of the particle positions. The probability density $\rho_1(x)$ for the measurement of the first particle is uniform as expected for system eigenstates. However, in the course of the measurements, $\rho_j(x)$ starts revealing a dark soliton notch and the last density $\rho_8(x)$ becomes hardly distinguishable from the mean-field (MF) dark soliton profile (black solid line) corresponding to the same average particle momentum as $P/N=\frac{\pi}{L}$ --- note that the mean-field profile has been shifted so that the soliton position $x_0$ matches the minimum of $\rho_8(x)$. (b) Phase of the mean-field solution (black solid line), whose probability density is shown in (a), and the phase of the wavefunction of the last particle (red dotted line) corresponding to the measurement realization presented in (a). Different simulations of the detection process lead to $\rho_8(x)$ with the notch localized at different points $x_0$. When one shifts the obtained particle positions so that $x_0$ always equals $\frac{L}{2}$, the average particle density can be prepared and it is shown in (c) (green stairs dash line) together with the mean-field dark soliton profile (black line). The average density bases on $10^4$ simulations of the measurement process. Deviations between the average density and the mean-field prediction seem to be due to small $N$ effects. That is, before the conditional probability density $\rho_j(x)$ converges to a final form, particles which are chosen according to different probability distributions distort the average density. Red stairs line in (c) shows the histogram where only positions of last particles are collected which agrees very well with the mean-field soliton profile. (d)-(f) are related to similar data as (a)-(c) but for the total momentum per particle $P/N=\frac{7\pi}{4L}$.}
\label{singlepar}
\end{figure}

Before we analyze results of simulations of the detection process let us describe what we can expect from the mean field description. For $L\rightarrow\infty$, the function $\sqrt{n}\tanh\left(\frac{x-x_0}{\xi}\right)$ represents a single dark soliton solution of the GPE with the notch localized at $x=x_0$. The size of the notch is determined by the healing length $\xi=\frac{1}{\sqrt{cn}}$. For $L<\infty$ and $\xi$ comparable with $L$, dark soliton profile is modified. Different soliton solutions, that fulfill periodic boundary conditions and are characterized by different average values of the particle momentum, are given in terms of the Jacobi functions \cite{carr00,kanamoto09, wu13}. There is a single-soliton solution with the average momentum that coincides with the momentum per particle for the $|sol\rangle$ eigenstate, i.e. with $P_{sol}/N=\frac{\pi}{L}$. The corresponding probability density is plotted in Fig.~\ref{singlepar}(a) together with a typical example of the conditional probability densities $\rho_j(x)$ obtained in the course of successive measurements of positions of particles that are prepared in the type~II eigenstate $|sol\rangle$ for $N=8$, $c=0.08$ and $L=1$, i.e. $\gamma=0.01$ (in this regime the mean field solution is not very different from a single particle solution). While $\rho_1(x)$ is uniform, the successive measurements reveal a clear notch in the probability densities. Typically, the measurement of the positions of 3 particles is sufficient to obtain $\rho_j(x)$ which changes little for the rest of the measurements and which is similar to the mean field dark soliton profile; see Fig.~\ref{singlepar}(a) --- note that the mean-field profile has been shifted so that the soliton position $x_0$ matches the minimum of $\rho_N(x)$. In Fig.~\ref{singlepar}(b) we show plots of the phase of the mean-field solution and the phase of the wavefunction of the last particle corresponding to the measurement realization presented in Fig.~\ref{singlepar}(a). The agreement is perfect and proves that the detection process drives the many-body state to the dark soliton wavefunction.

For $N=8$, a single simulation of the measurement sequence produces 8 particles positions only. In order to obtain an average particle density one has to perform many simulations. However, each simulation reveals the probability density $\rho_N(x)$, for the choice of the last particle, that is localized around a different point in the 1D space. From a plot of $\rho_N(x)$ we can infer at which $x_0$ the anticipated dark soliton is localized. Then, all results can be shifted so that $x_0$ always coincides with $\frac{L}{2}$ and the average particle density can be prepared as shown in Fig.~\ref{singlepar}(c). There are some deviations between the average density and the mean-field dark soliton profile which seem to be due to small $N$ effects. That is, in the course of the measurements, a minimum of $\rho_j(x)$ localizes as far as possible from points where the previous $(j-1)$ particles have been detected. Particles chosen before $\rho_j(x)$ converges to a final profile, have noticeable contribution to the average density if $N$ is as small as 8. In Fig.~\ref{singlepar}(c) we show also the average density of the last particles, i.e. those chosen according to the final $\rho_N(x)$, which nearly perfectly matches the mean-field dark soliton profile.

\begin{figure}
\includegraphics[width=0.9\columnwidth]{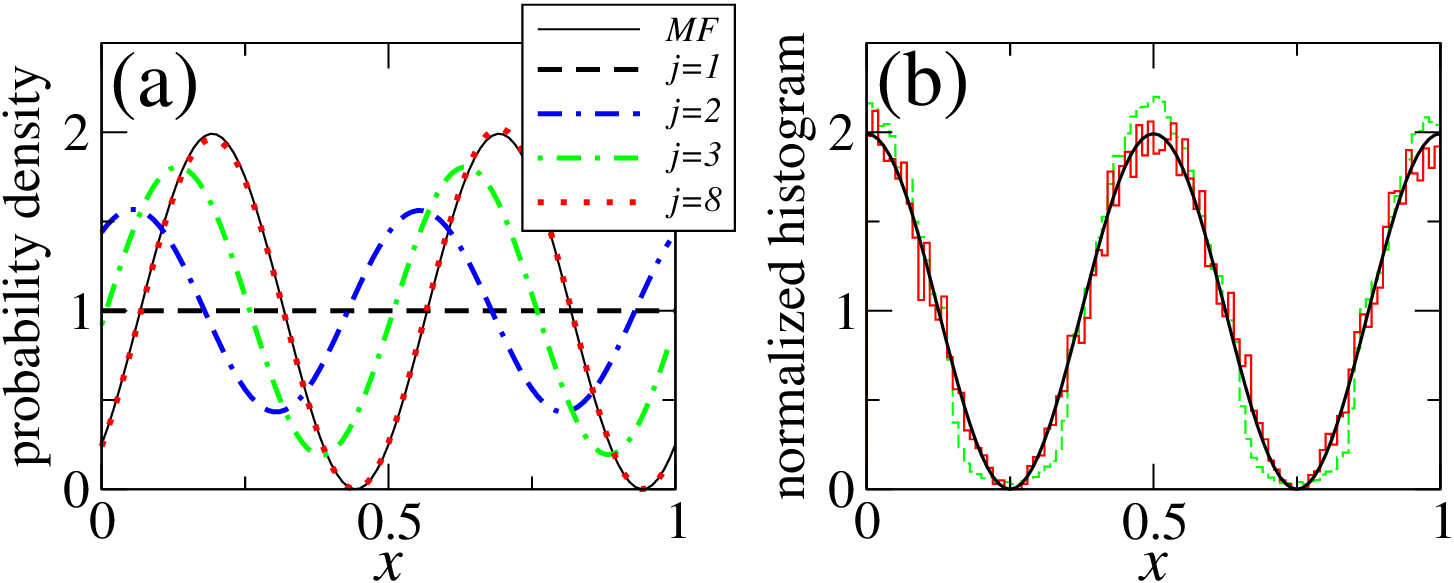}
\caption{(color online) Similar results as in Fig.~\ref{singlepar}(a) and Fig.~\ref{singlepar}(c) but for the eigenstate with the total momentum $P=0$ corresponding to two-hole excitations of the $N=8$ system with $c=0.08$ and $L=1$. Histograms in (b) have been prepared by shifting particle positions obtained in the simulations so that one  of the two minima of $\rho_N(x)$ always coincides with $\frac{L}{4}$.}
\label{double}
\end{figure}

For even $N$ we have analyzed the results of the measurements of the system prepared in the type~II eigenstate with the total momentum per particle $P/N=\frac{\pi}{L}$. There are other one-hole excitations corresponding to the momentum per particle $|P|/N\in\left\{\frac{2\pi}{LN},\frac{4\pi}{LN},\dots,\frac{2\pi}{L}\right\}$. Mean-field single-soliton solutions, with the average particle momentum in the same range, describe solitons which are rather {\it grey} than fully {\it black} because the density profiles do not reach zero at soliton positions. Especially the mean-field solution with the average particle momentum approaching $\frac{2\pi}{L}$ corresponds to the uniform density without any soliton signature, i.e. soliton solution degenerates into a uniform superflow state \cite{kanamoto08,kanamoto10}. In Fig.~\ref{singlepar}(d)-(f) we show the results of the particle position measurements performed in the $N=8$ system that is prepared in the type~II eigenstate with the total momentum per particle $P/N=\frac{7\pi}{4L}$. The mean-field approach predicts a {\it grey} soliton profile which agrees very well with the results of the simulations, especially with the average density of the last particles detected in each measurement sequence.

{\it Double soliton.}
We may suspect that creation of a two-hole excitation allows us to prepare the system in an eigenstate which is able to reveal double dark soliton profile in the course of the particle positions measurements. In Fig.~\ref{double} we show the results corresponding to such an eigenstate where in the sequence of half-integers $\{I\}_N$ that defines the ground state, we remove $-\frac12$ and $\frac12$ and add $-\frac{9}{2}$ and $\frac{9}{2}$. Then, the total momentum of the system remains zero as in the ground state. The results of the measurements agree very well with the mean-field double dark soliton profile corresponding to the vanishing average particle momentum. If one analyzes the phase of the wavefunction of the last particle in the measurement realization presented in Fig.~\ref{double}(a) it turns out it performs two discontinuous jumps by $\pi$ at the soliton positions and perfectly matches the behaviour of the mean-field phase.

\begin{figure}
\includegraphics[width=0.9\columnwidth]{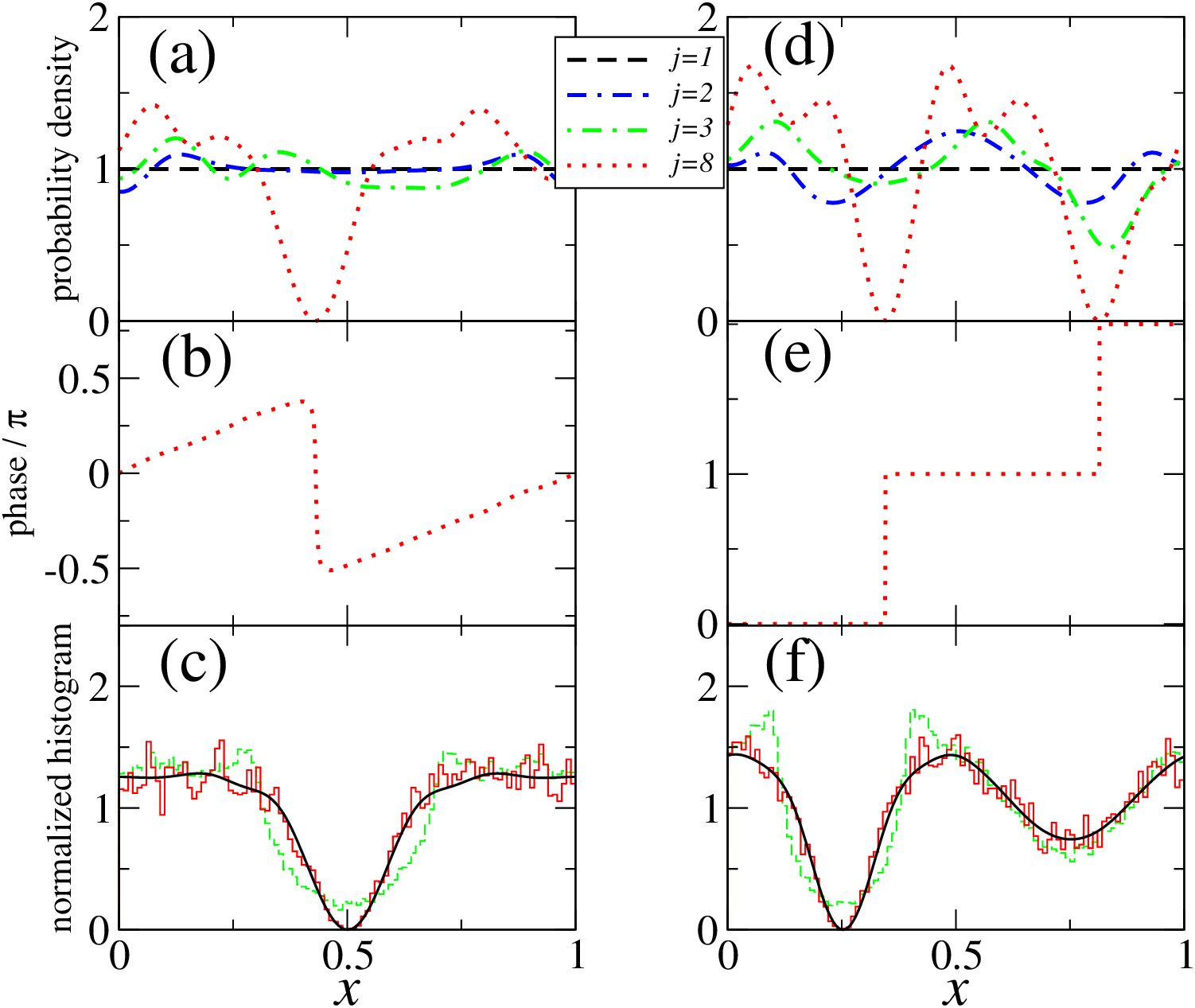}
\caption{(color online) Similar data as in Figs.~\ref{singlepar}-\ref{double} but for strong particle interactions, i.e. for $c=8$, $N=8$ and $L=1$. (a)-(c) correspond to the one-hole excitation with $P=\frac{N\pi}{L}$. (d)-(f) are related to the two-hole excitation with $P=0$. Histograms shown in (c) have been prepared by shifting particle positions obtained in the simulations so that the global  minimum of $\rho_N(x)$ always coincides with $\frac{L}{2}$. Histograms in (f) base on particle positions shifted so that one of the two minima of $\rho_N(x)$, as e.g. visible in (d), always coincides with $\frac{L}{4}$. Note that black lines in (c) and (f) are not related to mean-field prediction as in Figs.~\ref{singlepar}-\ref{double} but to $\rho_N(x)$ averaged over all (i.e. $10^4$) measurement realizations.}
\label{strong}
\end{figure}

{\it Strong interaction regime.}
So far we have investigated the system in the weak interaction limit, i.e. for $\gamma=0.01$. Now we switch to the strong interaction regime where the mean-field description is not valid. In this limit one- and two-hole excitations are defined in the same manner as we have done for the weak interactions. Examples of the particle positions measurements, for $\gamma=1$ (i.e. $c=8$, $N=8$), are presented in Fig.~\ref{strong} --- they correspond to the two eigenstates analyzed in the weak interaction regime. The conditional probability densities $\rho_j(x)$ are not as regular as for the weak interactions. In the case of the one-hole excitation corresponding to the total momentum $P=\frac{N\pi}{L}$, they converge to functions which show single minimum at a certain position $x_0$ where $\rho_N(x_0)\approx 0$. The phase of the wavefunction of the last particle in the measurement realization presented in Fig.~\ref{strong}(a) reveals jump at the notch position very similar as in the weak interaction case, cf. Fig.~\ref{singlepar}(b). In the case of the two-hole excitation with $P=0$, there are two pronounced minima of $\rho_N(x)$. The phase of the wavefunction of the last particle behaves very similarly as the phase of the double soliton solution analyzed in the weak interaction case, i.e. it performs jumps by $\pi$ at the minima of $\rho_N(x)$, see Fig.~\ref{strong}(e). Distance between the two minima is noticeably different in different realizations of the detection process. Thus, contrary to the weak interaction case, the relative distance between the two minima of $\rho_N(x)$ has probabilistic character and cannot be considered as a well defined classical quantity. When we prepare the average particle density shifting the obtained particle positions so that one of the two minima of $\rho_N(x)$ always coincides with $0.25L$, only one notch is clearly visible. The other notch, located around $x=0.75L$, is broad and shallow. The density notches visible in Figs.~\ref{strong}(c) and (f) are narrower than the notches in the weak interaction limit but much wider than expected from the corresponding healing length $\xi=\frac18$. Similar effect has been also observed in Ref.~\cite{sato12} where superposition of the type~II eigenstates allows the authors to create a notch in the reduced single particle probability density. 

In the numerical simulations it is necessary to truncate sums over system eigenstates in (\ref{rhoeq}) and (\ref{gam}). Strong interaction regime constitutes the most difficult case in the computer simulations. In this regime, for $N=8$ we have collected form factors $\langle\{k\}_{7}|\hat\psi(0)|\psi_0\rangle$ corresponding to ordered sets $\{I_1,\dots,I_7\}$ where $I_1\ge -9$  and $I_7\le 9$. Maximal values of the total momentum and quasimomentum of the resulting eigenstates are $5.7k_F$ and $6.2k_F$, respectively, where $k_F$ is the maximal quasimomentum related to the ground state of the system for $N=7$. Then, to calculate the reduced single particle density $\rho_1(x)$ we keep only those form factors whose absolute values are greater than $\alpha=10^{-3}$ times the maximal absolute value of the collected form factors. All these restrictions allows us to obtain integrals of the reduced single particle densities $\int_0^L\rho_1(x)dx$ which deviate from the unity less than $10^{-3}$. In the weak interaction limit, and for $\alpha=10^{-4}$, $\int_0^L\rho_1(x)dx$ equals 1 within the double precision.

In summary, we have considered the Lieb-Liniger model and concentrated on an analysis of the particle positions measurements. Eigenstates of the system related to hole excitations (the so-called type~II spectrum) are believed to correspond to dark soliton states in the weak interaction limit. We have shown that despite the fact that the corresponding reduced single particle probability densities are uniform, the measurements induce breaking of the translational symmetry and result in emergence of dark solitons which are localized at different positions in space in different realizations of the detection process. In the strong interaction regime, the results of the measurements also reveal signatures of notches in particle distributions. Their size is smaller than in the weak interaction case but greater than the healing length of the system. Interestingly, in the case of a two-hole excitation, two notches can be visible but their relative distance fluctuates from one realization of the detection process to another.

We are gratefull to Dominique Delande and Jacek Dziarmaga for valuable discussions.
Support of Polish National Science Centre via project number DEC-2012/04/A/ST2/00088 is acknowledged. The work was performed within the project of Polish-French bilateral programme POLONIUM and the FOCUS action of Faculty of Physics, Astronomy and Applied Computer Science of Jagiellonian University.

\end{document}